# POTENT AND MAX-FLOW ALGORITHMS

J.F.L. SIMMONS[1], A. NEWSAM[1], M.A. HENDRY[2]
[1] *University of Glasgow, Glasgow, UK*
[2] *University of Sussex, Falmer, Brighton, UK*

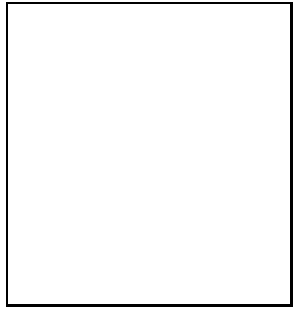

**Abstract**

Although POTENT purports to use only radial velocities in retrieving the potential velocity field of galaxies, the derivation of transverse components is implicit in the smoothing procedures. Thus the possibility of using nonradial line integrals to derive the velocity field arises. In the case of inhomogeneous distributions of galaxies, the optimal path for integration need not be radial, and can be obtained by using max-flow algorithms. In this paper we present the results of using Dijkstra's algorithm to obtain this optimal path and velocity field.

## 1 Introduction

One of the most effective methods for reconstructing peculiar velocity fields of galaxies on the scale of 100 Mpc from observed redshifts has been POTENT ([1], [3] hereafter DBF, [4]). The main weakness of this method is the sparseness of the data. Sky coverage of the most complete redshift surveys is still highly anisotropic. Some of this incompleteness will be overcome by more rigorous and deeper surveys. However the intrinsic distribution, clustering of galaxies and voids will always be a barrier to determining peculiar velocity fields accurately. In these circumstances it is important to find the optimal way of reconstructing the potential peculiar velocity field. This in turn could lead us to a much improved understanding of local density inhomogeneities, and further constraints on the value of $\Omega$ and on the various cold and hot dark matter model candidates.

POTENT has usually been presented as a method for obtaining the smoothed peculiar velocity field almost directly from observations of redshifts and distances of galaxies. Such an attractive idea often obscures the actual practice by which the smoothed potential field is constructed, and in so doing obscures also ways in which the method can be improved. In this paper we present a method for obtaining the velocity potential from *non-radial* paths. This opens up the possibility of choosing paths that pass through regions where the number density of galaxies is high, and where the errors of distance estimation are low. Such paths should yield a more accurate potential velocity field. To understand this new approach, we shall have to outline the main steps of the POTENT method of DBF.

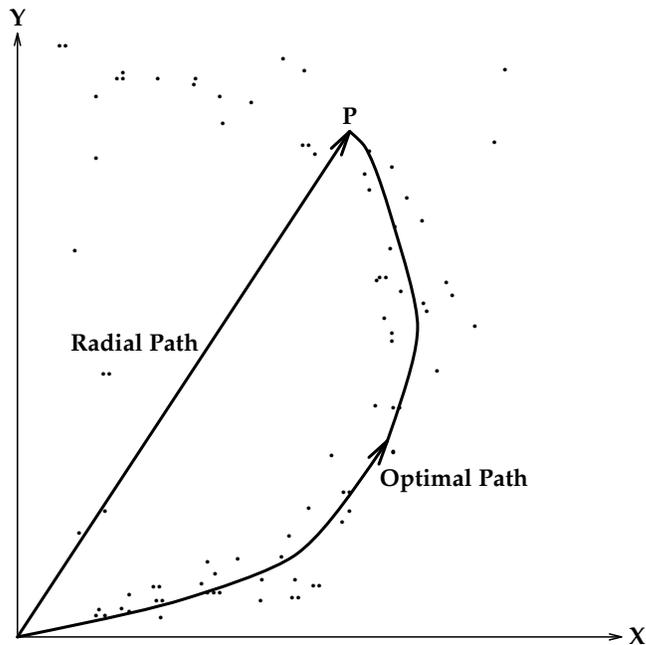

Figure 1: The optimal path minimises the error of the velocity potential, and will be pulled towards regions of high galaxy number density.

## 2  Orthodox Potent

The main steps of POTENT may be crudely summarised as the following

**1.** Take measured redshifts and distances of galaxies and hence obtain peculiar velocities of individual galaxies.

**2.** Construct an initial smoothed velocity field by using a tensor window function.

**3.** Take the radial component of this initial smoothed velocity field and carry out the line integral along a radial path to obtain the velocity potential.

**4.** Use the potential to derive the vectorial smoothed peculiar velocity field.

Of course there are numerous other considerations that are discussed elsewhere (DBF, [6]) but not crucial to our present discussion.

The smoothing procedure in step 2 furnishes us with a *vectorial* velocity field. DBF write down and use only the radial component of this field. Of course there are reasons to suppose the radial component will be more accurately determined than the transverse $(\theta, \phi)$ components since the redshifts are essentially telling us the radial components. However, even if the error on the radial component were a factor of ten smaller than on the transverse, it is still possible to gain advantage by taking non-radial paths. Figure 1 schematically depicts such a situation.

# 3  Errors on the Radial and Transverse Components

DBF obtain the initial smoothed velocity field, $\vec{v}(\vec{r})$ at arbitrary spatial point $\vec{r}$, by minimising

$$\sum_{i=1}^{n}(\vec{v}.\vec{e}_r(\vec{r}_i) - u_i)^2 W(\vec{r},\vec{r}_i) \tag{1}$$

where $\vec{r}_i$ is the position vector of the $i^{th}$ galaxy, and there are $n$ galaxies in the catalogue, $\vec{e}_r(\vec{r}_i)$ is the unit vector in the radial direction at the $i^{th}$ galaxy at position $\vec{r}_i$ and $W(\vec{r},\vec{r}_i)$ is the weighting or window function that determines the relative importance of the $i^{th}$ galaxy. Writing

$$\vec{v}(\vec{r}) = v_r \vec{e}_r(\vec{r}) + v_\theta \vec{e}_\theta(\vec{r}) + v_\phi \vec{e}_\phi(\vec{r}) \tag{2}$$

$$s^i_{rr} = \vec{e}_r(\vec{r}).\vec{e}_r(\vec{r}_i), \qquad s^i_{\phi r} = \vec{e}_\phi(\vec{r}).\vec{e}_r(\vec{r}_i), \qquad s^i_{\theta r} = \vec{e}_\theta(\vec{r}).\vec{e}_r(\vec{r}_i) \tag{3}$$

and for expediency $W^i = W(\vec{r},\vec{r}_i)$ minimisation yields

$$\sum_i W^i \begin{pmatrix} s^i_{rr}s^i_{rr} & s^i_{rr}s^i_{\theta r} & s^i_{rr}s^i_{\phi r} \\ s^i_{rr}s^i_{\theta r} & s^i_{\theta r}s^i_{\theta r} & s^i_{\theta r}s^i_{\phi r} \\ s^i_{rr}s^i_{\phi r} & s^i_{\theta r}s^i_{\phi r} & s^i_{\phi r}s^i_{\phi r} \end{pmatrix} \begin{pmatrix} v_r \\ v_\theta \\ v_\phi \end{pmatrix} = \sum_i \begin{pmatrix} u_i s^i_{rr} W^i \\ u_i s^i_{\theta r} W^i \\ u_i s^i_{\phi r} W^i \end{pmatrix} \tag{4}$$

which can be written

$$\mathbf{AV} = \mathbf{b} \tag{5}$$

in accordance with the notation of DBF (appendix A). $\mathbf{V}$ is simply the column matrix of components of the velocity. Evidently the inversion of this equation yields all three components of the initial smoothed velocity field.

Since the estimated distances of galaxies are subject to error there will be a corresponding error on $\vec{v}(\vec{r})$. Let us write the initial smooth peculiar velocity field obtained from one realisation as

$$\hat{\vec{v}}(\vec{r}) = \vec{v}(\vec{r}) + \delta\hat{\vec{v}}(\vec{r}) \tag{6}$$

where the hat indicates an *estimated* velocity. DBF's (appendix A) linear error analysis in which they derive a bias and variance on the radial component $v_r$ of $\vec{v}(\vec{r})$ can easily be generalised to the vectorial case, although we shall not do this here. The estimated components $\hat{v}_r(\vec{r})$, $\hat{v}_\phi(\vec{r})$, and $\hat{v}_\theta(\vec{r})$ are not statistically independent. If we assume that the bias has been removed from $\hat{v}_r(\vec{r})$ so that $E(\delta\hat{v}_r(\vec{r})) = 0$, we can write the covariance as

$$E(\delta\hat{v}_i(\vec{r})\delta\hat{v}_j(\vec{s})) = R_{ij}(\vec{r},\vec{s}) \tag{7}$$

$E$ denotes expected value. The velocity autocorrelation function, $R_{ij}(\vec{r},\vec{s})$, will depend on the window function, number density, the dispersion of the distance estimator and also to some extent on the input peculiar velocities. Our main purpose in this paper is to demonstrate the viability of the method and we shall not attempt to accurately model $R_{ij}(\vec{r},\vec{s})$.

# 4  Optimal Paths

The potential, $\Phi(\vec{r})$, of the peculiar velocity field is given by the path integral

$$\Phi(\vec{r}) = -\oint_0^{\vec{r}} \vec{v}(\vec{s}).d\vec{s} \tag{8}$$

The error, $\delta\hat{\Phi}(\vec{r})$ arises only from the error, $\delta\hat{\vec{v}}(\vec{s})$ in the estimated initial peculiar velocity since the path is prescribed. It is therefore given by

$$\delta\hat{\Phi}(\vec{r}) = -\oint_0^{\vec{r}} \delta\hat{\vec{v}}(\vec{s}).d\vec{s} \qquad (9)$$

The optimal path for obtaining the potential of the velocity field at position $\vec{r}$ we shall take to be that path for which the variance of the line integral is minimum.

If we assume that $\hat{\vec{v}}$ is unbiased so that $E(\delta\hat{\vec{v}}) = 0$, then evidently

$$E(\delta\hat{\Phi}) = 0 \qquad (10)$$

The variance of $\delta\hat{\Phi}$ is given by

$$E(\delta\hat{\Phi}(\vec{r}))^2 = E\left(\oint_0^{\vec{r}} \delta\hat{\vec{v}}(\vec{s}).d\vec{s} \oint_0^{\vec{r}} \delta\hat{\vec{v}}(\vec{t}).d\vec{t}\right) = \int_0^{\mu} \int_0^{\nu} R_{ij}(\vec{s}(\mu),\vec{t}(\nu))x^{i'}(\mu)x^{j'}(\nu)d\mu d\nu \qquad (11)$$

where primes denote derivatives with respect to the arguments. Thus the optimal path will be given by

$$\delta \int_0^{\mu} \int_0^{\nu} R_{ij}(\vec{s}(\mu),\vec{t}(\nu))x^{i'}(\mu)x^{j'}(\nu)d\mu d\nu = 0 \qquad (12)$$

In the case where the autocorrelation function is a delta function, i.e.

$$R_{ij}(\vec{s},\vec{t}) = \delta^3(\vec{s}-\vec{t})\sigma_{ij}(\vec{s}) \qquad (13)$$

equation (12) simply defines a geodesic on a riemannian space with $\sigma_{ij}$ as its metric tensor. Generally, however, we can expect components of the initial smoothed field at different spatial points to be correlated, and the correlation length to be of the same order of magnitude as the effective radius of the window function.

Although equation (12) is very interesting from a mathematical viewpoint, we shall not proceed further along those lines. In practice we do not know the form of the autocorrelation function, and it could be best approximated by numerical simulations. A more natural way to proceed is to use finite element methods, which we now discuss.

## 5 Dijkstra's Algorithm

We wish to calculate the 'best' velocity potential, from which the peculiar velocity field may be obtained by taking the gradient. This requires determining the potential $\Phi$ at regular grid points, at least in regions of space where the galaxies are sufficiently dense for the reconstruction to be meaningful. Suppose we have $N$ gridpoints at which we wish to evaluate the potential of the velocity field.

Let us assume that the error in going between the $a^{th}$ and the $b^{th}$ gridpoint is prescribed. This 'error length', which we shall call an arcweight, should depend on the number density of galaxies in the joint neighborhood of both gridpoints, the distance between the gridpoints and the distance of both from the origin. We take the potential to be zero at the first gridpoint (our galaxy), and so wish to find the path along which the total error is least. At first sight it might appear that this is an *NP-complete* problem. Luckily this is not the case. Dijkstra's algorithm (cf [7]) finds the exact solution to this problem in a number of steps that is bounded above by $N^2$. This algorithm is one of a class of MAXFLOW algorithms used in network theory.

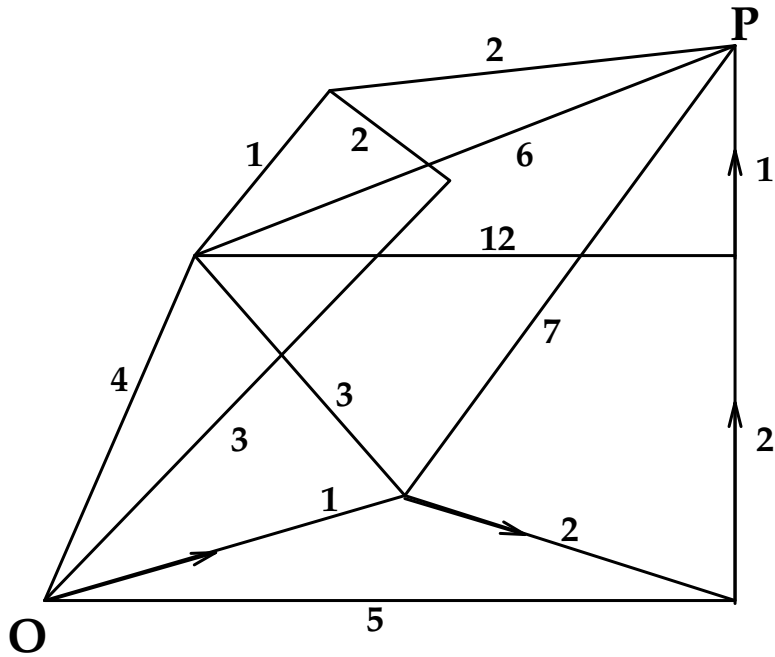

Figure 2: Shortest path from O to P, indicated by arrows, can be obtained using Dijkstra's algorithm.

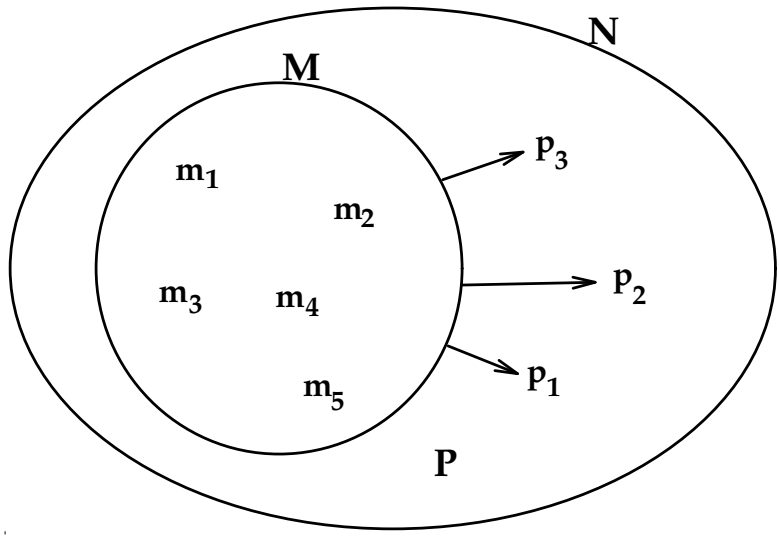

Figure 3: Illustration of Dijkstra's algorithm. **M** is set of nodes for which optimal paths from $m_1$ have been found. **N** is the set of all nodes and $\mathbf{P} = \mathbf{N} - \mathbf{M}$.

## 5.1 Description of Dijkstra's Algorithm

Figure 2 gives a schematic representations of a network of nodes (gridpoints). We write the set of nodes as

$$\mathbf{N} = \{n_1, n_2, n_3, ... n_N\} \tag{14}$$

The arcweight between each pair of nodes has been calculated according to some prescription. Let the arcweight between node $n_a$ and node $n_b$ be $c_{ab}$. We shall assume that $c_{ab} = c_{ba}$, although this is not strictly necessary for what follows. A path will be determined by its nodes. Thus the path $n_{a_1} n_{a_2} n_{a_3} n_{a_4} ... n_{a_M}$ has $M$ nodes and total arcweight or pathlength

$$c_{a_1 a_2 a_3 .. a_M} = c_{a_1 a_2} + c_{a_2 a_3} + ... c_{a_{M-1} a_M} \tag{15}$$

Clearly we may also write

$$c_{a_1 a_2 a_3 .. a_M} = c_{a_1 a_2 a_3 ... a_{M-1}} + c_{a_{M-1} a_M} \tag{16}$$

The problem is to find the path from node $n_1$ to every other node $n_a$ that minimises the sum of arcweights. We proceed as follows.

Start from $n_1$. Rename $n_1$ as $m_1$ (see Figure 3). Suppose at any stage we have a set $\mathbf{M}$ of nodes to which we have established the minimum pathlength. Let the remaining nodes form the set $\mathbf{P}$. Define $s_\mathbf{M}(p)$ to be the shortest pathlength from $m_1$ to $p$ *that uses only intermediate nodes* in $\mathbf{M}$. This must also be the shortest path from $m_1$ to $p$. Now choose from the set of nodes $\mathbf{P}$ the node $p$ for which the path length is shortest, and add it to the set $\mathbf{M}$ and repeat until until the set $\mathbf{M}$ contains all the nodes.

## 6 Application of Dijkstra's algorithm to Potent and Results

Our main problem in applying this method to POTENT is how to establish the arcweights between nodes. The arguments we present now are largely heuristic. Errors between two nodes (gridpoints) we expect to be determined largely by the number density of galaxies in the mutual neighbourhood of the two nodes. The higher the galaxy number density the smaller the error. Radial components of the initial peculiar velocity field will probably have lower errors than the transverse components, but we might expect the two transverse components to have the same errors. If we take the gridpoints $n$ and $m$ to be separated by more than the correlation length of the autocorrelation function, then we can assume that $\delta \vec{v}(m)$ and $\delta \vec{v}(n)$ are uncorrelated. Hence it would be reasonable to take

$$E(\delta \hat{\Phi}(\vec{r}))^2 = \sum E(\delta \vec{v}(ab) . \Delta \vec{x}(ab))^2 = \sum E(\delta v_i(ab) \delta v_j(ab)) \Delta x^i(ab) \Delta x^j(ab) \tag{17}$$

where $n_a$ and $n_b$ are consecutive nodes along the path of integration, $\Delta \vec{x}(ab)$ the separation between these two gridpoints, and $\delta v_i(ab)$ is the error in the $i^{th}$ component of the initial smoothed peculiar velocity evaluated at the midpoint of the segment. Since the variance of the distance estimator increases with radial distance squared, we shall take the arcweight to also scale with $r^2$. To simplify, we disallow arcs between gridpoints more than three gridlengths away, and assume that

$$\sigma_{r\theta} = \sigma_{r\phi} = \sigma_{\phi\theta} = 0 \text{ and } \sigma_{rr} = k^2 \sigma_{\theta\theta} = k^2 \sigma_{\phi\phi} \tag{18}$$

where $k$ is some parameter.

Thus we shall take $c_{ab}$ to be of the form

$$c_{ab} = n^\alpha r^2 ((\Delta r)^2 + k^2 r^2 (\sin^2 \theta (\Delta \phi)^2 + (\Delta \theta)^2) \tag{19}$$

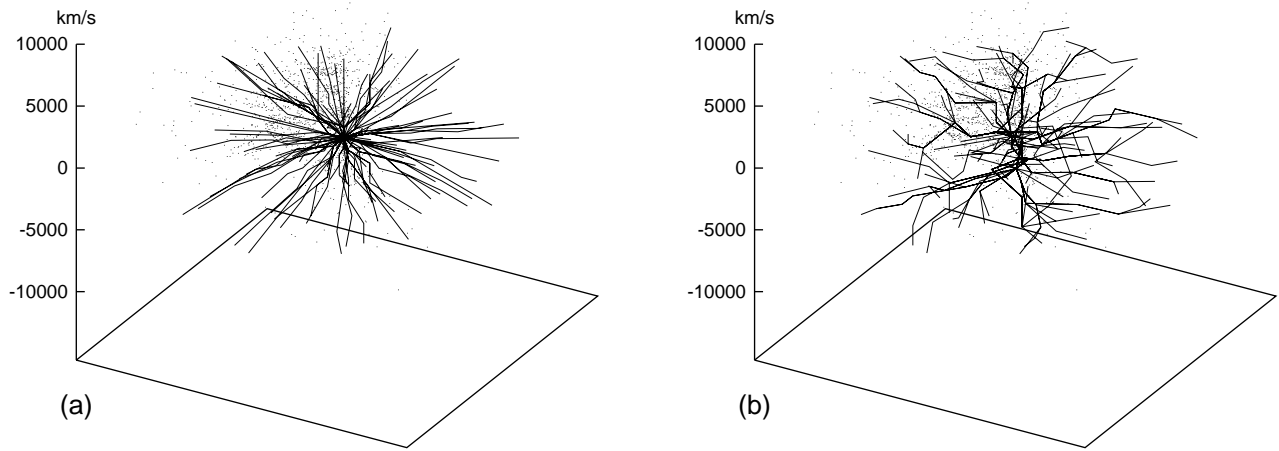

Figure 4: Some optimal paths with number density of galaxies indicated by dots. Paths in figure (a) have low density weighting and most are almost radial. In figure (b), paths have high density weighting and are non-radial.

where $n$ denotes the number density of galaxies. The value of $k$ in the above equation essentially tells one the errors on the transverse components of the initial field compared with the radial. This will obviously depend on the window function. It will also depend on the actual peculiar velocity field. Very rough simulations indicate that for some simple peculiar velocity field the transverse components will be poorly recovered, and consequently $k$ will be large, typically between 5 and 10.

We have taken a spatial distribution of galaxies to be given by [5] combined with [2] data. Two peculiar velocity fields are taken corresponding to quiet Hubble flow and uniform streaming. Galaxy distances are subjected to distance errors, and the minimum length paths found using arcweights of the form (19) to define the line integral (8), and hence the velocity potential. Figures 4 and 5 show the optimal paths for two different heuristic arcweight functions, and their corresponding rederived velocity fields.

For both fields it turns out that the optimal paths are almost radial. By insisting on arcweights that are heavily dependent on density, and for which $k \sim 1$ one can achieve non-radial paths. However, in these cases the recovered potential velocity field is noisy and bears little resemblance to the input field.

## 7 Conclusion and Discussion

The density inhomogeneities are not great enough to produce optimal paths that are highly non-radial when realistic arcweight functions are chosen. Although in regions where the data is dense paths deviating from radial do produce potential velocity fields in agreement with the radial paths there seems little advantage can be obtained in this way. The reasons for this somewhat disappointing result lies primarily with the large radius of the window function which tends to smooth out the effect of number density inhomogeneities. For slowly varying peculiar velocity fields the transverse components of the initial smoothed peculiar velocity obtained from using the window function will be poorly determined, except perhaps near to our own galaxy, where in any case the recovery from radial paths is good. Improvements in the derived potential velocity could be made by integrating along differentiable curves rather than along rectilinear line segments, but probably the gains would not be commensurate with the effort

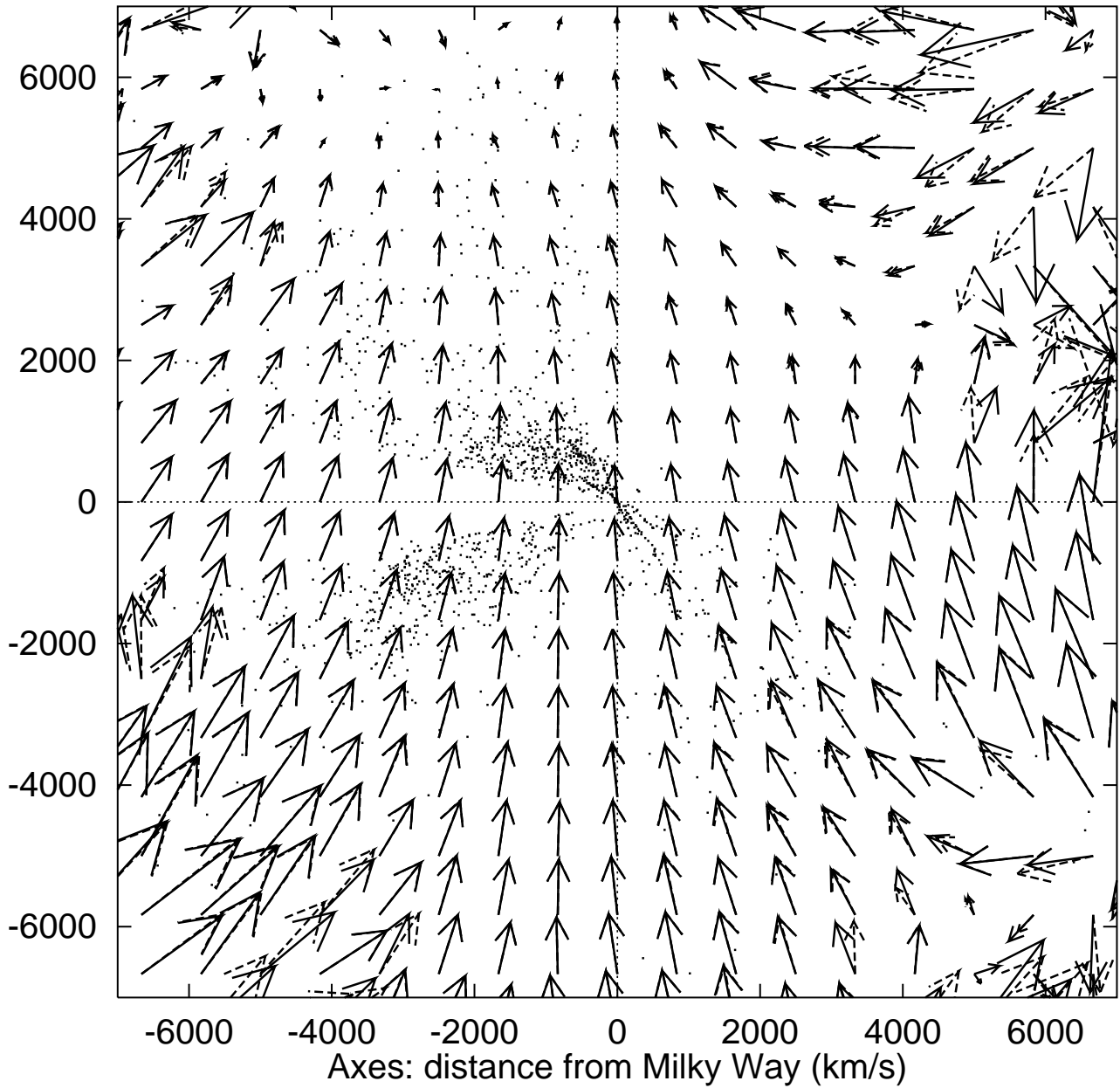

Figure 5: Potential velocities derived from optimal paths obtained for different density weightings ($\alpha$) and ratios of transverse to radial errors ($k$). The solid arrows are the velocities for $\alpha = 0$ and $k = 9$. Dashed arrows for $\alpha = -4$ and $k = 5$. Galaxy number density is projected onto the plane.

involved.

One way to improve on the derived potential would be to take the potential at each gridpoint averaged over many paths, which could be weighted according to their total arcweight. The weakness of POTENT seems to us largely to stem from the arbitrary choice of window function. The notion of using an ensemble of paths which are not necessarily radial to obtain an ensemble averaged potential is one that can transcend the use of ad hoc window functions, and possibly overcome the attendant difficulties of sample gradient and Malmquist bias. We are currently investigating this possibility.

**Acknowledgements.** The authors would like to thank Dr. Lewis Mackenzie of Glasgow University Department of Computing Science for drawing Dijkstra's algorithm to their attention.